\documentclass[twocolumn,showpacs,preprintnumbers,amsmath,amssymb,aps,prl]{revtex4}
\usepackage{graphicx}% Include figure files
\usepackage{dcolumn}% Align table columns on decimal point
\usepackage{bm}% bold math

\newcommand{\B}{{\cal B}}
\newcommand{\Bb}{{\overline\B}}

\newcommand{\Ct}{{\widetilde C}}
\newcommand{\I}{{\cal I}} 
\newcommand{\Ib}{{\overline\I}} 
\newcommand{\N}{{\cal N}}
\newcommand{\M}{{\cal M}}
\newcommand{\Mb}{{\overline\M}} 
\renewcommand{\P}{{\mathbb P}}
\newcommand{\R}{{\mathbb R}}
\newcommand{\Z}{{\mathbb Z}}
\renewcommand{\Im}{\mathrm{Im}}
\renewcommand{\Re}{\mathrm{Re}}

\begin{document}
\title{A semiclassical realization of infrared renormalons}
\author{Philip Argyres }
\affiliation{Physics Dept., U.\ of Cincinnati, Cincinnati OH 45221}
\author{Mithat \"Unsal}
\affiliation{Dept.  of Physics and Astronomy, SFSU, San Francisco, CA 94132}
\affiliation{SLAC and Department of Physics, Stanford University, CA 94025}

\date{\today}
\begin{abstract}
Perturbation series in quantum field theory are generally divergent  asymptotic series which are also typically not Borel resummable in the sense that the resummed series is ambiguous.  The ambiguity is associated with singularities in the Borel plane on the positive real axis.  In quantum mechanics there are cases in which the ambiguity that arises in perturbation theory cancels against a similarly ambiguous contribution from instanton--anti-instanton events.  In asymptotically free gauge theories this mechanism does not suffice because perturbation theory develops ambiguities associated with singularities in the Borel plane which are closer to the origin by a factor of about $N$ (the rank of the gauge group) compared to the singularities realized by instanton events.  These are called IR renormalon poles, and on $\R^4$ they do not possess any known semiclassical realization.  By using continuity on $\R^3 \times S^1$, and by generalizing the works of Bogomolny and Zinn-Justin to QFT, we identify saddle point field configurations, e.g., bion--anti-bion events, corresponding to singularities in the Borel plane which are of order $N$ times closer to the origin than the 4d BPST instanton--anti-instanton singularity.  We conjecture that these are the leading singularities in the Borel plane and that they are the incarnation of the elusive renormalons in the weak coupling regime.  
%This set of ideas may lead to a non-perturbative continuum definition of certain QFTs. 
\end{abstract}
\maketitle

Perturbation theory in almost all interesting quantum field theories is divergent even after proper regularization and renormalization.  A method that defines a finite perturbative expansion is to resum it by first computing its convergent Borel transform as a function of a complex parameter, $t$, and then appropriately integrating the Borel transform along the positive  real axis in the $t$-plane.  But in most cases the perturbative expansion is not Borel resummable due to certain types of singularities of the Borel transform.  These obstructions amount to an ambiguity in the would-be Borel sum.  Therefore, if we take the Borel procedure as the definition of  perturbation theory, perturbation theory by itself is ill-defined.

In quantum mechanics there are cases, e.g., the double-well potential, in which this disease of the Borel sum can be cured 
by a procedure that we refer to as the Bogomolny--Zinn-Justin (BZJ) prescription.  Here the perturbative ambiguity cancels against a non-perturbative contribution from instanton--anti-instanton events \cite{Bogomolny:1980ur, ZinnJustin:1981dx}.  The sum of the perturbative and non-perturbative semiclassical expansions in quantum mechanics apparently produces ambiguity-free (and accurate) results \cite{ZinnJustin:2004ib}.  But it proved impossible to generalize this idea to asymptotically free field theories like QCD because of the occurrence of \emph{infrared (IR) renormalon} singularities.  These singularities are much closer to the origin of the Borel $t$-plane than the BPST instanton--anti-instanton singularities \cite{'tHooft:1977am, Beneke:1998ui}, and there are no known semiclassical configurations against which to cancel the IR renormalon ambiguities.  

In this work, by combining the ideas of \cite{Bogomolny:1980ur, ZinnJustin:1981dx} and \cite{Unsal:2007vu, Unsal:2007jx} involving compactification, continuity, and semiclassical analysis, we argue that the appropriate sum of perturbative and non-perturbative semiclassical expansions can be used to give a non-perturbative continuum definition of a large class of gauge field theories.

\vspace{-5mm} 
\subsection{Borel resummation and the BZJ prescription in quantum mechanics}

We first review the Borel resummation idea, which we take as the definition of perturbation theory.  Let $P(g^2) = \sum_{n=0}^\infty a_n\, g^{2n}$ denote a perturbation series for an observable.  If $P(g^2)$ has a convergent Borel transform $BP(t) := \sum_{n=0}^\infty a_n\, t^{n}/n!$ for positive real $t$, then 
\begin{equation}\label{PP}
\P(g^2) =  \frac{1}{g^2} \int_0^\infty BP(t)\, e^{-t/g^2} dt 
\end{equation}
formally gives back $P(g^2)$ and we say $\P(g^2)$ is the (unique) Borel resummation of $P(g^2)$.  However, if $BP(t)$ has singularities at $t_i \in \R^+$, then $\P(g^2)$ as defined by the integral  (\ref{PP}) is \emph{ambiguous}.

The ambiguity in $\P(g^2)$ can be seen as the freedom to choose integration contours $C_\pm$ in (\ref{PP}) to go just above or just below the singularity.  (Equivalently, one can keep the contour on the real axis but move the singularity location by shifting $g^2 \rightarrow g^2 \mp i \epsilon$.)  Define
\begin{equation}\label{PPpm}
\P_\pm(g^2) :=  \frac{1}{g^2}\int_{C_\pm} BP(t)\, e^{-t/g^2} dt .
\end{equation}
Then $\P_\pm(g^2) = \Re\P(g^2) \pm i \Im\P(g^2)$, where $\Im\P(g^2) = e^{-t_1/g^2 } + e^{-t_2/g^2 } + \ldots$. 
 
An equivalent way to describe this ambiguity is to consider $\P(g^2)$ as a function in the complex $g^2$-plane.  The ambiguity of $\P(g^2)$ for $g^2>0$ is a branch cut in the complex $g^2$-plane, and $\P(g^2)$ defines an analytic function there except at the cut.  In particular, very often a perturbative sum $P(g^2)$ that is non-Borel summable for $g^2>0$ is Borel summable for negative $g^2$, where the answer is unique.  However, the analytic continuation from negative to positive $g^2$ is ambiguous since it can be done either in a clockwise sense, $\Ct_+$, or in a counter-clockwise sense, $\Ct_-$, ending at $g^2 \pm i \epsilon$.  The Borel sums obtained in this way by continuations $\Ct_\pm$ in the $g^2$-plane are equivalent to the Borel sums in (\ref{PPpm}) with choices of $t$-plane contour $C_\mp$, respectively.

Borel sum ambiguities of the type $\pm i e^{-t_1/g^2 }$ should be viewed as a defect of perturbation theory.  But it need not be a problem in the full theory, and may actually provide a link between the perturbative and non-perturbative physics.  First, the ambiguity in $\P(g^2)$ has the same form as an instanton contribution: $e^{-t_1/g^2} \sim e^{-2 S_I}$ where $S_I$ is the instanton action.  Furthermore, though there is no ambiguity associated with an instanton amplitude $[\I]$ and it cannot cancel the ambiguity of perturbation theory, an \emph{instanton--anti-instanton} amplitude $[\I\Ib]$, on the other hand, does have a two-fold ambiguity \cite{Bogomolny:1980ur, ZinnJustin:1981dx}.  The identification of this ambiguity and its cancellation against the ambiguity of the perturbative Borel sum is what we call the BZJ prescription.
   
Let us review this argument as we will find conditions under which it also occurs in field theory.  While the space-time position of a single instanton is an exact zero mode, there is a relatively weak attractive interaction between an $\I$-$\Ib$ pair, making their relative separation a quasi-zero mode, parametrically split from the other non-zero modes.  One integrates over this quasi-zero mode in evaluating the $[\I\Ib]$ amplitude.

Since the interaction is attractive, the integral over the quasi-zero mode is dominated by small separations where a well-defined $\I$-$\Ib$ configuration does not exist, rendering the amplitude meaningless.  But since $\I$-$\Ib$ configurations carry the same quantum numbers as the perturbative vacuum, one must treat them as one does in perturbation theory where one takes $g^2$ negative to make it Borel summable.  When $g^2<0$ the $\I$-$\Ib$ interaction is repulsive at short distances and the quasi-zero mode integral is concentrated at some separation, $r$, much larger than the instanton size, but much smaller than the typical single instanton separation \cite{Unsal:2012zj}.  Hence, such a defect should be considered as a molecular instanton and at distances much larger than $r$ it can be treated as being point-like.  The quasi-zero mode integral converges and the $[\I\Ib]$ amplitude, obtained by analytically continuing back to positive $g^2$, is two-fold ambiguous depending on the choice of the continuation path, $\Ct_\pm$.  We call these two amplitudes $[\I\Ib]_\pm$, respectively.  As asserted above, the continuation of $\P(g^2)$  from $g^2<0$ to $g^2>0$ also produces a two-fold ambiguity.  The ambiguity in the sum of the Borel summed and bi-instanton amplitudes vanishes,
\begin{equation}
\Im[ \P_\pm(g^2) + [\I\Ib]_\pm (g^2) ] = 0, 
\label{cancel}
\end{equation}
up to terms of order $e^{-t_2/g^2} \ll e^{-t_1/g^2}$ \cite{ZinnJustin:1981dx, ZinnJustin:2004ib}.  This means that the ambiguities at order $e^{-t_1/g^2}$ cancel, independent of the choice of path, so long as one consistently uses $C_\mp$ along with $\Ct_\pm$.
   
Thus although perturbation theory by itself is not meaningful, the sum of both the perturbative and non-perturbative parts of the semiclassical expansion does seem to be meaningful.  This is the essence of the BZJ prescription.

Can this idea  work in field theory, e.g., in QCD?  In \cite{'tHooft:1977am} 't Hooft argued that it does not work for gauge theories on $\R^4$ due to the above-mentioned IR renormalon problem.  We will argue that it does work on $\R^3 \times S^1$ in a gauge theory continuously connected to one on $\R^4$, thus providing a new perspective on 't Hooft's renormalons.

\vspace{-5mm} 
\subsection{QCD on $\R^4$: instantons and renormalons}

On $\R^4$, in a QCD-like gauge theory with gauge group $G$, the instanton--anti-instanton amplitude calculated in the same way as above (for small 4d instantons for which one can do a semiclassical analysis) gives a contribution $[\I_4\Ib_4]_\pm \sim  \pm i e^{-2n S_I}$ \cite{Bogomolny:1977ty}.  Therefore, according to an argument of Lipatov \cite{Lipatov:1976ny} correlating the large-order behavior of perturbation series with certain saddle point configurations, the Borel transform has singularities at 
\begin{eqnarray}
t=t_n= n (2S_I)g^2 = 16\pi^2 n, \qquad 0<n\in\Z,
\end{eqnarray}
leading to $\Im \P_{\pm}(g^2) \approx \pm i e^{-t_n/g^2}= \pm i e^{-2n S_I}$.  The $[\I_4\Ib_4]_\pm$ singularities, as in the case of quantum mechanics, arise due to the $n!$ growth in the number of Feynman diagrams, and the two ambiguities cancel.  But this is far from a happy ending.

The Borel transform $BP(t)$ has other (far more important) singularities closer to the origin of the Borel plane, located at   \cite{'tHooft:1977am, Beneke:1998ui}. 
\begin{equation}
t = \tilde t_n = n (2S_I)g^2/\beta_0 , \; \;   n=2,3, \ldots
\label{renormalons}
\end{equation}
where $\beta_0 \sim N$ is the first coefficient of the beta function and $N$=rank$(G)$.  These come from the leading divergence of perturbation theory due to the sub-class of ``bubble" diagrams (and not due to the $n!$ growth in the total number of Feynman diagrams).  This class of diagrams grow as $(n/2)!$ at $n$-th order in perturbation theory with the main contribution coming from small internal momenta of order $\Lambda$, the strong coupling scale of QCD, and give the poles in the Borel plane located at (\ref{renormalons}).  't Hooft called these singularities ``IR renormalons" in the expectation/hope that they would be shown to be associated with a semiclassical saddle-point-like instanton.  However, no such configuration is known to date.

Thus the disease of perturbation theory is not cured in field theory as it is in the quantum mechanics examples.  This is not just a formal problem, but a reflection of a basic and troubling lack of understanding of gauge theories.  It raises the conceptual question of whether a continuum definition of these theories exists \cite{'tHooft:1977am}.  More practically, in QCD phenomenology ``power-law corrections" are invoked to remove the renormalon pole ambiguity in  $\P(g^2)$ \cite{Beneke:1998ui} but without a microscopically justified and concrete method to compute them.

\vspace{-5mm} 
\subsection{QCD(adj) on $\R^3 \times S^1$: topological molecules}

A new program to study 4d gauge dynamics is to use compactification on $\R^3 \times S^1$ where $S^1$ is, crucially, a spatial (non-thermal) circle with size $L$.  This amounts to using periodic (not anti-periodic)  boundary conditions for fermions.  With this compactification, a class of gauge theories exists which have no center-symmetry changing phase transition or no phase transition at all as the radius is varied, in contradistinction to the thermal case.  Such theories exhibit  semiclassical calculability at fixed $N$  and small $L$ \cite{Unsal:2007vu, Unsal:2007jx, Anber:2011de},  and volume independence  in the large-$N$ limit  \cite{Kovtun:2007py}.  A gauge theory in this category is Yang-Mills theory with $n_f$ adjoint Majorana (or Weyl) fermions, abbreviated as QCD(adj). 

On a small $S^1$, the dynamics of QCD(adj) is weakly coupled.  The gauge holonomy around the $S^1$ behaves as a compact adjoint Higgs field, and the gauge group abelianizes at long distances,  $G \rightarrow U(1)^N$, where $N = {\rm rank}(G)$, similar to the Coulomb branch of $\N{=}2$ supersymmetric theories \cite{Seiberg:1996nz} and to the 3d Polyakov model \cite{Polyakov:1976fu}.  The low energy theory is a collection of 3d compact $U(1)$'s with fermions.  Although the long-distance theory is 3d, the fact that the microscopic theory lives in 4d is crucial for confinement and other non-perturbative properties \cite{Unsal:2007jx}.  

Monopole-instantons $\M_i$, $i=1, \ldots, N+1$, contribute at leading non-perturbative order in the semiclassical expansion.  These are associated with the simple roots and the affine root of $G$ \cite{Lee:1997vp,Kraan:1998sn}.  Each $\M_i$ carries $2n_f$ fermion zero modes.  So, unlike in the Polyakov mechanism in 3d, they do not induce a mass gap and confinement 
\cite{Affleck:1982as,Unsal:2007jx}

The 4d BPST instanton carries $2n_f h^\vee$ fermion zero modes where $h^\vee \sim N$ is the dual Coxeter number of $G$.  The $\M_i$'s may be viewed as $h^\vee$ constituents of the 4d BPST instanton.  The action of these monopole events is $S_{\M_i} \sim S_{\I_4}/N$.  Therefore 4d BPST instantons are relatively unimportant in the semiclassical expansion since they only appear at about $N$-th order.

At second order in the semiclassical expansion there are two types of topological molecules which lead to amplitudes with no fermionic zero modes.  These are in one-to-one correspondence with the non-vanishing elements of the extended Cartan matrix $\hat A_{ij}$ of $G$. 
%\begin{itemize}
%\item 

For each entry $\hat A_{ij}<0$ there exists a \emph{magnetic bion}, $\B_{ij} \sim [\M_i\Mb_j]$, a topological molecule with no fermionic zero mode but with a net magnetic charge.  The $\B_{ij}$ generate a mass gap for gauge fluctuations and confinement in QCD(adj) \cite{Unsal:2007jx}.
%\item 

For each diagonal entry, $\hat A_{ii}>0$, there exists a \emph{neutral bion}, $\B_{ii} \sim  [\M_i\Mb_i]$, with zero topological charge and zero magnetic charge.  The $\B_{ii}$ generate a repulsion among the eigenvalues of the gauge holonomy \cite{AU}.
%\end{itemize}

In QCD(adj), neither the contribution of the magnetic bions nor of the neutral bions to amplitudes is ambiguous.  One might have suspected that the neutral bion would give an ambiguous contribution as it has the same (vanishing) quantum numbers as the perturbative vacuum.  The fact that it is not so in QCD(adj) is tied to the fermionic zero modes of its constituents. 
In a  purely bosonic theory,  the neutral bion  $\B_{ii}$ is the leading   effect (\emph{second} order in the semiclassical expansion)  to generate a non-perturbative ambiguity, see below. 
  That instanton--anti-instanton amplitudes can be  ambiguity-free in a theory with fermions, but are (necessarily) ambiguous in a purely bosonic theory 
is already known in the context of quantum mechanics   \cite{Balitsky:1985in}.   A classification of the effects of various types of saddle point configurations will be given in \cite{AU}.

\vspace{-5mm} 
\subsection{The bion--anti-bion as renormalon}

In QCD(adj), we concentrate on the contribution at \emph{fourth} order in the semiclassical expansion due to (quasi)-saddle point configurations associated to magnetic bions and magnetic anti-bions,  $[\B_{ij}\Bb_{ij}]$ molecules.   They experience a Coulomb attraction, and are indistinguishable from the perturbative vacuum, according to their total topological and magnetic charge.   We therefore have to compute the contribution of these correlated overall-neutral bion pair  amplitudes in the same way we do the purely perturbative contributions, following the natural generalization of the BZJ prescription to QFT.  

Therefore, first take $g^2$ negative. Then the bion constituents repel at small separations and the integral over the quasi-zero mode is finite and concentrated around some characteristic separation.  In the integral over the quasi-zero mode, one also needs to subtract the effect of uncorrelated bions which were already taken into account at second order in the semiclassical expansion.  This step is the same as in quantum mechanics \cite{Bogomolny:1980ur, ZinnJustin:1981dx}.  Now analytically continue back to positive $g^2$.  In doing so, we find \cite{AU} a two-fold ambiguity in the $[\B_{ij}\Bb_{ij}]$  amplitude, namely, $[\B_{ij}\Bb_{ij}]_\pm = \Re[\B_{ij}\Bb_{ij}]  \pm i \Im[\B_{ij}\Bb_{ij}]$ where $\Im[\B_{ij}\Bb_{ij}] \approx \exp^{-4S_{\M_i}}$.  Then, following Lipatov \cite{Lipatov:1976ny}, we predict poles in the Borel plane located at 
\begin{equation}
t_n= 4n S_{\M_i} g^2 \approx 4n  S_{\I_4}/N , \qquad 0<n\in\Z. 
\end{equation}

So we have found semiclassical saddle point configurations for QCD(adj) on $\R^3 \times S^1$ giving Borel plane singularities on the order of $N$ times closer to the origin than the 4d BPST instanton--anti-instanton singularity.  This is the same neighborhood as the IR renormalon singularities of 't Hooft, at least parametrically in $N$.  In order for weakly coupled continuum QCD(adj) on $\R^3 \times S^1$ to make sense, we must have
\begin{align}\label{borelbion}
\Im \P_{\pm}  +   \Im  [\B_{ij}  \Bb_{ij}] _{\pm}=0 \quad \text{for QCD(adj).}
\end{align}
We make two conjectures: 
\begin{itemize}
\item[{\bf 1)}] The same set of bubble diagrams in 4d which give the IR renormalon singularity also give the $[\B_{ij}\Bb_{ij}]$ singularity in perturbation theory around the abelianized $U(1)^N$ vacuum.
\item[{\bf 2)}] Abelianizing gauge theories with two-index representation fermions on $\R^3 \times S^1$ have no other singularities closer to the Borel plane origin. 
\end{itemize}
If these two conjectures are true, then these theories may be non-perturbatively defined in the continuum through their semiclassical expansions. 

\vspace{-5mm}  
\subsection{General gauge theories}
 
These arguments can be generalized to other gauge theories.
In $\N{=}4$ and $\N{=}2$ supersymmetric extensions of pure 
Yang-Mills theory,  despite the fact that $\M_i$'s exist, neither the neutral bion nor the magnetic bion does \cite{Poppitz:2011wy}.  The simplest way to see this is to observe that the number of fermion zero modes of the $\M_i$'s prohibit a superpotential and, consequently, a bosonic potential.  This means that on $\R^3 \times S^1$ these theories have no singularities on the positive real axis in the Borel plane, are therefore Borel summable at finite $S^1$, and, by continuity and analyticity, on $\R^4$. This argument is complementary to and in agreement with that of Ref.\ \cite{Russo:2012kj}. 

In pure Yang-Mills, since there is a phase transition on $\R^3 \times S^1$ as the radius is reduced, one might conclude that this formalism does not apply.  However, there exists a smooth continuation of the large-$S^1$ confined phase to a small-$S^1$ (weakly coupled) confined phase.  The small-$S^1$ theory obtained in this manner is called deformed Yang-Mills (dYM) theory \cite{Unsal:2008ch}.  In dYM the neutral bion $\B_{ii} = [\M_i\Mb_i]$ has an ambiguity corresponding to a pole in the Borel plane located at $t_n= 2n S_{\M_i} g^2 = 2n  S_{\I_4}/N$.  This is again in the same neighborhood, parametrically in $N$, as the IR renormalon poles in YM on $\R^4$ \cite{'tHooft:1977am}.  In this case the neutral bion is the semiclassical incarnation of the IR renormalon.  We expect an ambiguity-free definition of dYM due to cancellations such as 
\begin{align}\label{borelbion2}
\Im\P_\pm + \Im[\M_i\Mb_i] _\pm=0  \quad \text{for dYM.}
\end{align}

A speculation in \cite{'tHooft:1977am} is that IR renormalon singularities might somehow be related to quark confinement.  Indeed, in the semiclassical regime we see that the proliferation of $\B_{ij}$ events in QCD(adj), and the proliferation of $\M_i$ events in dYM  generate a mass gap and confinement  \cite{Unsal:2007jx,Unsal:2008ch}.  So, at least in the semiclassical domain, a sharpening of 't Hooft's speculation is that the events at order $n$ in the semiclassical expansion which cause confinement are responsible at order $2n$ for generating IR renormalon singularities.

Finally, we note that the pole locations associated with our topological molecules do differ from the proposed IR renormalon pole locations on $\R^4$ by numerical (order one) factors.  This difference stems from the fact that our analysis is in the semiclassical $LN \Lambda \lesssim 1$ domain. 
 As we increase $L$  or $N$, we expect these poles to saturate to their values on $\R^4$ in the volume independence $LN\Lambda \gg1$ domain. 
   This last observation, in principle, should permit us to study IR renormalons through equivalent  large-$N$ matrix models 
 \cite{Kovtun:2007py, Bringoltz:2009kb}, combined with  the techniques of  Refs.~\cite{Gross:1982at, Marino:2007te}.  

\begin{acknowledgments}
It is a pleasure to thank E. Poppitz for helpful discussions and comments. 
\end{acknowledgments}

\bibliography{QCD}% Produces the bibliography via BibTeX.

\end{document}